\documentclass{article}
\usepackage{cite}
\usepackage{graphicx}
\usepackage{amsmath}
\usepackage{amsfonts}

\begin{document}

\title{Comment on 'Optimization of photonic crystal structures'}

\author{Andreas H\aa kansson and Jos\'{e} S\'{a}nchez-Dehesa}

\author{Wave Phenomena Group, Nanophotonic Technology Center, \\
Polytechnic University of Valencia,\\
C/ Camino de Vera s/n, E-46022, Valencia, Spain} 

\maketitle 

\begin{abstract}
Recently, Smajic \textit{et al.} \cite{Smajic} published an article on numerical structural optimizations of two-dimensional photonic crystals using two different classes of optimization algorithms i.e, deterministic for local searches and stochastic for global. In this comment we reexamine some of their conclusions regarding the global stochastic search strategies. It is concluded that the steps taken to increase the efficiency of the optimization of the test problem chosen by Smajic \textit{et al.}, which was selected to keep the search space tractable, can be misleading when applied to intractable problems.
\end{abstract}

\section{Introduction}
First of all, when evaluating the performance of an optimization strategy the class of optimization problems at hand play a very crucial part. In 1997 Wolpert and Macready \cite{Wolpert} published the 'No free lunch theorems for optimization'. The theorems proof that for any algorithm, any elevated performance over one class of problem is exactly paid for in bad performance over another class. In other words, it does not exist no better nor worse optimization algorithm with respect to its average performance on all possible classes of problems. In Ref. \cite{Smajic}, Smajic \textit{et al.} analyze the performance of four different global search strategies by optimizing a power divider coded by 12 binary parameters. The 12 parameters give rise to a total of 4096 possible configuration. The relatively small size of the problem was chosen to keep the search space tractable, and by doing so, they limit themself to a specific class of problems. If this is not done carefully this class might be inadequately represented in a global perspective and the method will lack generalization. 

\section{An example}
As a counter example the problem of optimizing a doped PC lens has been chosen. The number of lattice sites has been increased to 182, which results in a considerable increase in the size of the search space. Now a problem with $2 ^{182}=6.1 \times 10^{54}$ possible configurations has to be processed. Because of the immense number of solutions it is impossible to do a brute-force evaluation of all these configurations making the optimization process indispensable. Please refer to Ref. \cite{Sanchis} for any further details about this specific problem. To design the lens structure a design tool based on a multiple scattering theory \cite{Ishimaru}, as a direct solver, and a genetic algorithm (GA) \cite{Goldberg} for global optimization, was implemented. This approach is similar to one of the four global search algorithms analyzed in section 4 of Ref. \cite{Smajic}. The authors of Ref. \cite{Smajic} was "disillusioned" by the poor performance of the standard GA for solving a 12 bit-parameters problem of designing a power divider implementing the GA with a population of 45 individuals. As a countermeasure they used a micro-GA that works with multiple runs of a standard GA implemented with a much smaller population. In this way they improved the performance of the optimization process considerably. Here we have repeated the some of the steps taken by Smajic \textit{et al.} and shown that there conclusions lack the generality of the concept "\textit{Optimization of photonic crystal structures}". 

\begin{figure}[h]
	\centering
		\includegraphics[width=1.0\textwidth]{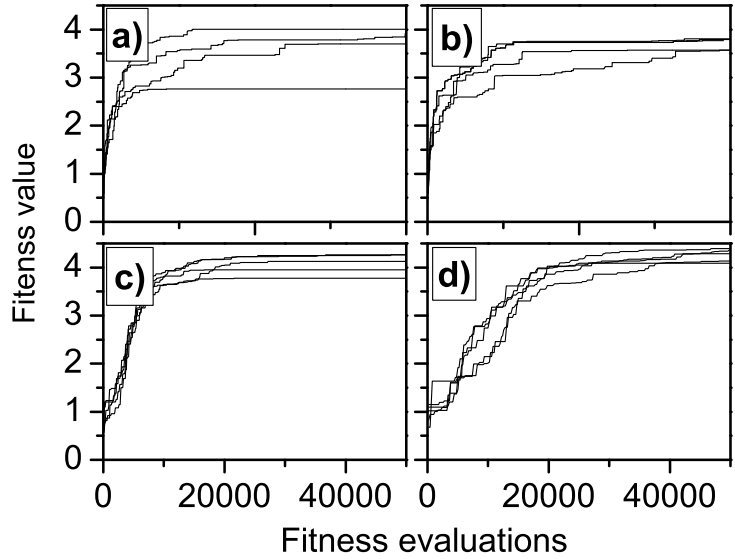}
	\caption{GA popultion dependecy. The best fitness in the population at every generation is printed out for four different GAs. The four GAs differ in population sizes and are applied to a problem coded with 182 bit-parameters. The following population sized were used 5, 20, 100 and 200 in a), b), c) and d) respectively. Each GA is executed 5 times corresponding to the five lines in each graph. The average fitness value after 50'000 evaluations and for 5 independent runs are for a) 3.34, b) 3.69, c) 4.06 and for d) 4.25}
	\label{fig:4graphs}
\end{figure}

Figure \ref{fig:4graphs} picture the performance of different GAs implemented to optimize the 182 bit-parameters lens problem. The GAs are implemented with different population sizes. Figure \ref{fig:4graphs}a and b correspond to the so called micro-GA with population sizes of 5 and 20, respectively, while Fig. \ref{fig:4graphs}c and d correspond to normal GA optimizations with populations of 100 and 200 respectively. Each optimization was carried out five times to get an idea of their stochastic behavior and each run was stopped after 50'000 evaluations. It is clearly demonstrated that for this particular problem, a so called micro-GA (fig \ref{fig:4graphs}a and b) performs bad and a standard GA with a large population (fig \ref{fig:4graphs}b and c) is considerable more robust and effective. These results contradict the conclusions drawn in Ref. \cite{Smajic} from the 12-bit power divider design problem.

\section{Conclusions}
In summary, the optimization or inverse-design of photonic crystal structures is an challenging problem. To find a global search algorithm for solving the wide class of problems consisting in effective design of PC devices is a difficult task that has to be tackled with rigor. To analyze the performance in a generic way of any proposed scheme of optimization it is important to choose a general problem or even to solve a set of problems. In our opinion, this has not been done in Ref.  \cite{Smajic} and, therfore, their conclusions can be misleading.

\newpage
\section*{List of Figure Captions}

Fig. 1. GA popultion dependecy. The best fitness in the population at every generation is printed out for four different GAs. The four GAs differ in population sizes and are applied to a problem coded with 182 bit-parameters. The following population sized were used 5, 20, 100 and 200 in a), b), c) and d) respectively. Each GA is executed 5 times corresponding to the five lines in each graph. The average fitness value after 50'000 evaluations and for 5 independent runs are for a) 3.34, b) 3.69, c) 4.06 and for d) 4.25
\pagebreak

\end{document}